\documentclass[11pt,superscriptaddress,aps,prd,preprint]{revtex4}
\usepackage[active]{srcltx}
\usepackage{graphicx}
\usepackage[utf8]{inputenc}
\usepackage{amsmath}
\usepackage{xcolor}
\usepackage{times,txfonts}
\usepackage{float}

\begin{document}

\title{Relativistic Bose-Einstein condensate in the rainbow gravity}

\author{J. Furtado}
\affiliation{Centro de Ci\^{e}ncias e Tecnologia, Universidade Federal do Cariri, 63048-080, Juazeiro do Norte, Cear\'{a}, Brazil}
\email{job.furtado@ufca.edu.br}

\author{J. F. Assun\c{c}\~{a}o}
\affiliation{Departamento de f\'{i}sica, Universidade Federal do Maranh\~{a}o, 65080-805, S\~{a}o Lu\'{i}s, Maranh\~{a}o, Brazil}

\author{C. R. Muniz}
\affiliation{Faculdade de Educa\c{c}\~{a}o, Ci\^{e}ncias e Letras de Iguatu, Universidade Estadual do Cear\'{a}, 63-500-000, Iguatu, Cear\'{a}, Brazil}
\email{celio.muniz@uece.br}

\begin{abstract}

In this paper, we study the effects of a modified theory of gravity - the rainbow gravity - on the relativistic Bose-Einstein condensate (BEC). We initially discuss some formal aspects of the model in order to compute the corrections to the relevant quantities of the condensate. Following, we evaluate the generating functional from which obtain some thermodynamic parameters. Then we calculate the corrected critical temperature $T_c$ that sets the relativistic Bose-Einstein condensate considering the three principal rainbow functions, finding, in addition, a phenomenological upper bound for the parameters involved in the model. Finally, we discuss how harder is for the particles at an arbitrary temperature $T<T_c$ to enter the condensed state compared to the usual scenario, {\it i.e.}, without rainbow gravity.

\end{abstract}

\maketitle

\section{Introduction}

Some models of quantum gravity presuppose that there exists an invariant energy scale (or length) associated with Planck's one \cite{Amelino1,Amelino2,Magueijo,Galan}, which are known as Double Special Relativity (DSR). In the DSR, the modified dispersion relation naturally leads to the Planck-scale departure from the Lorentz symmetry. According to these theoretical proposals, introducing quantum corrections to the Einstein-Hilbert action is not the only possible consistent modification to the General Relativity.

The approaches to quantum gravity suggesting that the usual energy-momentum dispersion relation becomes deformed near the Planck scale also predict that the spacetime is affected. Thus, its geometry changes according to the energy of the test particle in it. The mentioned approaches receive the generic name of rainbow's gravity, being studied in several scenarios such as string field theory \cite{Samuel}, loop quantum gravity \cite{Pullin}, and non-commutative geometry \cite{Carroll}. Some proposals combine both the approaches (e.g., corrections to both the action and dispersion relation), as in \cite{Channuie}. Thus, a nonlinear representation of the Lorentz transformations in momentum space would yield an energy-dependent spacetime geometry, leading to the mentioned modification of the relativistic energy-momentum dispersion relation \cite{Camelia2}, in a type of spacetime backreaction.

Such a semi-classical approach allows explaining, for instance, the currently observed ultra-high-energy cosmic rays, whose origin is still unknown, suggesting that the dispersion relation is in fact modified, which opens up new possibilities for theoretical developments. In astrophysics, for example, the study of the influence of the rainbow's gravity on the black hole properties, including their thermodynamics, was carried out in the most different scenarios \cite{Leiva,Li,Ali,Valdir1,Feng,Li1,Ronco,Ednaldo,Dehghani,Valdir3,Hamil}, as well as in the research on cosmic strings \cite{Valdir1.5,Bakke,Santos}. Such a modified theory of gravity can also play a decisive role in cosmology, specifically in the understanding of the early Universe, in which the involved energies are close to the Planck scale, contributing thus to avoid the initial singularity \cite{Awad,Majumder,Valdir2,Nozari,Hendi,Marc}.

The Bose-Einstein condensate (BEC) can have been present in the primordial Universe \cite{Takeshi} at some critical redshift parameter $z$. It can even be identified to the early dark matter, if this was constituted by bosons, with its perturbations possibly leaving an imprint in CMB (\cite{Hermano}, and references therein). Therefore, in such cosmological scenarios, it is quite reasonable to admit that the early BEC was under the influence of the rainbow gravity, as well as in actual high energy events.

In the context of high-energy physics, the BEC was recently studied in connection with Lorentz symmetry violation \cite{Furtado:2020olp, Casana:2011bv, Tian:2021mka}. It was shown in \cite{Furtado:2020olp, Casana:2011bv} how the CPT-even and CPT-odd Lorentz violating terms modify the BEC critical temperature and the thermodynamic parameters. In \cite{Furtado:2021tzd} the influence of the anisotropic scaling in the BEC was discussed. The possibility of probing low-energy Lorentz violating effects in dipolar BEC was addressed in \cite{Tian:2021mka}. Also, the BEC in Rindler space and its occurrence due to Unruh effect was studied in \cite{Takeuchi:2015nga}.

In this paper, we study the effects of rainbow gravity on low-energy phenomena, namely, on BEC, by considering corrections to some thermodynamics quantities associated to it, discussing some formal aspects of the model first. The calculations were carried out by computing the generating functional, from which we extract those quantities. The corrected critical temperature $T_c$ that sets the Bose-Einstein condensate was also computed for the three mostly adopted cases (functions) of the rainbow gravity. Moreover, we have obtained an upper bound for a combination of the parameters involved in the model from experimental measurements. Finally, we have discussed how harder is for the particles at an arbitrary temperature $T<T_c$ to pertain to the condensate state when compared with the usual scenario without rainbow gravity.

The paper is organized as follows: in the next section we present a brief review on the rainbow gravity. In section III we discuss the model, i.e., the complex scalar sector modified by the rainbow functions. We also compute the thermodynamic parameters, such as pressure, energy, specific heat at constant volume and charge density. The BEC critical temperature was calculated for three configurations of the rainbow functions separately in the the subsections $A$, $B$ and $C$. In section IV we present our final remarks.

\section{Rainbow gravity brief review}

The rainbow gravity was first studied in the context of Double Special Relativity (DSR) and it emerges as a generalization to curved spacetime of the deformed Lorentz symmetry group (locally). One of its consequences is the arising of a modified energy–momentum dispersion relation. Such modification is usually written in the form \cite{Magueijo, Magueijo:2002am, Magueijo:2002xx}
\begin{equation}
    E^2f^2(\epsilon)-p^2c^2g^2(\epsilon)=m^2c^4,
\end{equation}
where $f(\epsilon)$ and $g(\epsilon)$ are the so called rainbow functions, and $\epsilon=E/E_P$ being $E$ the energy of the probe particle and $E_P$ the Planck energy. In the low-energy limit the rainbow functions converges to unit, restoring the standard dispersion relation. However in the high-energy limit the rainbow functions end up violating the usual energy-momentum dispersion relation. The modification of this latter corresponds to a change in the metric, according to \cite{Magueijo:2002xx}, so that the Minkowski spacetime becomes
\begin{equation}
    ds^2=\frac{dt^2}{f^2(\epsilon)}-\frac{1}{g^2(\epsilon)}\delta_{ij}dx^idx^j.
\end{equation}

In order to study the rainbow gravity effects on the Friedmann-Robertson-Walker (FRW) universe \cite{Awad:2013nxa, Ali:2014xqa}, the following rainbow functions were considered
\begin{eqnarray}
    f(\epsilon=E/E_P)=1,\,\,\,g(\epsilon)=\sqrt{1-\xi (E/E_P)^s},
\end{eqnarray}
where $s>1$ and $\xi$ is a dimensionless free parameter of the model, which we will consider the same as the other rainbow functions to facilitate comparison between the employed models.

Another interesting choice for the rainbow functions is the following,
\begin{eqnarray}
    f(\epsilon=E/E_P)=g(\epsilon=E/E_P)=\frac{1}{1-\xi(E/E_P)}.
\end{eqnarray}
Such rainbow functions were considered in \cite{Magueijo, Magueijo:2002am} (and references therein) in studying possible nonsingular universe solutions and in \cite{Magueijo:2002xx}, since it assures a constant light velocity, it may provides a solution for the horizon problem.

A last choice of rainbow functions of great interest is given by
\begin{eqnarray}
    f(\epsilon=E/E_P)=\frac{e^{\xi(E/E_P)}-1}{\xi(E/E_P)},\,\,\, g(\epsilon=E/E_P)=1.
\end{eqnarray}
This choice of the rainbow functions was originally considered in \cite{Camelia2} in the context of Gamma Ray Bursts. Later, this same choice was also addressed in \cite{Awad:2013nxa, Santos:2015sva} in connection with FRW solutions.

\section{Model}\label{casez2}
The model we are considering consists of rainbow gravity extension of the complex scalar sector. Hence the lagrangian describing the system is
\begin{eqnarray}\label{lagrangian1}
\mathcal{L}&=&f^2(\epsilon)(\partial_0\phi)^{*}(\partial_0\phi)-g^2(\epsilon)(\partial_i\phi)^{*}(\partial_i\phi)-m^2\phi^{*}\phi.
\end{eqnarray}

The lagrangian (\ref{lagrangian1}) possesses a clear $U(1)$ symmetry, so that
\begin{equation}
    \phi\rightarrow \phi'=e^{-i\alpha}\phi,
\end{equation}
with $\alpha\in\mathbb{R}$. From Noether's theorem it is known that for any given continuous symmetry there is a conserved quantity in connection. To find out such conserved quantity let us consider $\alpha=\alpha(x)$, i.e., a spacetime position-dependent function. The Euler-Lagrange equation gives us the equation of motion for the ``field'' $\alpha(x)$. Since the contribution $\partial\mathcal{L}/\partial\alpha=0$, we can find the following charge density
\begin{eqnarray}
    \nonumber Q&=&\int d^3x j^0\\
    &=&\int d^3x\left[if^2(\epsilon)\left(\phi^{*}\frac{\partial\phi}{\partial t}-\phi\frac{\partial\phi^{*}}{\partial t}\right)\right].
\end{eqnarray}

The equations of motion for $\phi$ and $\phi^{*}$, given by
\begin{eqnarray}
-f^2(\epsilon)\partial_0^2\phi+g^2(\epsilon)\partial_i^2\phi-m^2\phi&=&0\\
-f^2(\epsilon)\partial_0^2\phi^{*}+g^2(\epsilon)\partial_i^2\phi^{*}-m^2\phi^{*}&=&0,
\end{eqnarray}
shows us, by a straightforward calculation, the conservation of the four current, i.e., $\partial_{\mu}j^{\mu}(x)=0$. Splitting the fieds $\phi$ and $\phi^{*}$ into two real components $\phi_1$ and $\phi_2$ as
\begin{eqnarray}
    \phi&=&\frac{1}{\sqrt{2}}(\phi_1+i\phi_2)\\
    \phi^{*}&=&\frac{1}{\sqrt{2}}(\phi_1-i\phi_2),
\end{eqnarray}
we can rewrite the Lagrangian, more conveniently, in terms of $\phi_a$ with $a=1,2$ as follows
\begin{eqnarray}
    \mathcal{L}&=&\frac{f^2(\epsilon)}{2}\partial_{0}\phi_a\partial_{0}\phi_a-\frac{g^2(\epsilon)}{2}\partial_{i}\phi_a\partial_{i}\phi_a-\frac{m^2}{2}\phi_a\phi_a.
\end{eqnarray}
The canonically conjugated momenta are:
\begin{eqnarray}\label{momenta}
    \pi_a=f^2(\epsilon)\partial_0\phi_a
\end{eqnarray}
then the Hamiltonian becomes
\begin{eqnarray}
    \mathcal{H}=\frac{1}{2}\left[\frac{1}{f^2(\epsilon)}\pi_a\pi_a+g^2(\epsilon)(\vec{\nabla}\phi_a)\cdot(\vec{\nabla}\phi_a)+m^2\phi_a\phi_a\right].
\end{eqnarray}
We can also express the charge density in terms of $\phi_a$,
\begin{equation}\label{chargedensity1}
    Q=\int d^3x\epsilon_{ab}\pi_a\phi_b.
\end{equation}
Letting $\mathcal{H}(\phi,\pi)\rightarrow\mathcal{H}(\phi,\pi)-\mu\mathcal{N}(\phi,\pi)$, where $\mathcal{N}(\phi,\pi)$ is the conserved charge density, identified as $Q$, and $\mu$ is the chemical potential, the partition function in the grand canonical ensemble becomes:
\begin{eqnarray}\label{Z1}
    Z=\int D\pi_a\int D\phi_a\exp\left\{\int_0^{\beta}d\tau\int d^3x\left[i\pi_a\frac{\partial\phi_a}{\partial\tau}-\mathcal{H}(\phi_a,\pi_a)+\mu\epsilon_{ab}\pi_a\phi_b\right]\right\}.
\end{eqnarray}
The closed functional integral sign stands for the fact that the field is constrained in such way that $\phi(\vec{x},0)=\phi(\vec{x},\beta)$ with $\beta=1/T$. The partition function can be written as
\begin{eqnarray}\label{Z2}
    Z=\int D\pi_a\int D\phi_a\exp\left\{\int_0^{\beta}d\tau\int d^3x\left[-\frac{1}{2f^2(\epsilon)}\pi_a^2+\left(i\frac{\partial\phi_a}{\partial\tau}-\mu\epsilon_{ab}\phi_b\right)\pi_a-\frac{g^2(\epsilon)}{2}(\vec{\nabla}\phi_a)^2-\frac{m^2}{2}\phi_a^2\right]\right\}.\,\,\,
\end{eqnarray}

The integration over the momenta can be directly done. Then we obtain,
\begin{eqnarray}\label{Z3}
    Z=(N')^2\int D\phi_a\exp\left\{\int_0^{\beta}d\tau\int d^3x\left[\frac{f^2(\epsilon)}{2}\left(i\frac{\partial\phi_b}{\partial\tau}-\mu\epsilon_{ab}\phi_a\right)^2-\frac{g^2(\epsilon)}{2}(\vec{\nabla}\phi_a)^2-\frac{m^2}{2}\phi_a^2\right]\right\}.
\end{eqnarray}
The factor $N'$ is a normalization constant, but since multiplication of $Z$ by any constant does not change the thermodynamics the factor $N'$ is irrelevant. The components of $\phi$ can be Fourier-expanded as,
\begin{eqnarray}\label{Fourier}
    \phi_1&=&\sqrt{2}\zeta\cos\chi+\sqrt{\frac{\beta}{V}}\sum_{n}\sum_{\vec{p}}e^{i(\vec{p}\cdot\vec{x}+\omega_n\tau)}\phi_{1;n}(\vec{p})\\
    \phi_2&=&\sqrt{2}\zeta\sin\chi+\sqrt{\frac{\beta}{V}}\sum_{n}\sum_{\vec{p}}e^{i(\vec{p}\cdot\vec{x}+\omega_n\tau)}\phi_{2;n}(\vec{p}),
\end{eqnarray}
where $\omega_n=2\pi n T$, owing to the constraint of periodicity that $\phi(\vec{x},\beta)=\phi(\vec{x},0)$ for all $\vec{x}$. Here $\zeta$ and $\chi$ are spacetime position independent parameters and determine the full infrared behaviour of the field; that is, $\phi_{1;0}(\vec{p}=\vec{0})=\phi_{2;0}(\vec{p}=\vec{0})=0$. This allows for the possibility of condensation of the bosons into the zero-momentum state. Substituting (\ref{Fourier}) into (\ref{Z2}) the partition function becomes
\begin{equation}
    Z=(N')^2\prod_n\prod_{\vec{p}}\int D\phi_{1;n}(\vec{p})D\phi_{2;n}(\vec{p})e^{S},
\end{equation}
being $S$ is given by

\begin{eqnarray}
     S=\beta V(f^2(\epsilon)\mu^2-m^2)\zeta^2-\frac{1}{2}\sum_n\sum_{\vec{p}}\left(\phi_{1;-n}(-\vec{p}),\phi_{2;-n}(-\vec{p})\right)D\left(\begin{array}{c}
\phi_{1;n}(\vec{p})   \\
\phi_{1;n}(\vec{p})
    \end{array}\right),
\end{eqnarray}
where $D$

\begin{eqnarray}
    D=\beta^2\left(\begin{array}{cc}
    f^2(\epsilon)\omega_n^2+\omega^2-f^2(\epsilon)\mu^2 & -2f^2(\epsilon)\mu\omega_n \\
    2f^2(\epsilon)\mu\omega_n & f^2(\epsilon)\omega_n^2+\omega^2-f^2(\epsilon)\mu^2
    \end{array}\right),
\end{eqnarray}
with $\omega=\sqrt{g^2(\epsilon)\vec{p}^2+m^2}$. Performing the integrations over $\phi_{1;n}$ and $\phi_{2;n}$, we have,
\begin{equation}
    \ln Z=\beta V[f^2(\epsilon)\mu^2-m^2]\zeta^2+\ln (\det D)^{-1/2},
\end{equation}
so that we can rewritte in the following form
\begin{eqnarray}\label{generatingfunctional}
    \ln Z=\beta V\left[f^2(\epsilon)\mu^2-m^2\right]\zeta^2-V\int\frac{d^3p}{(2\pi)^3}\left\{\frac{\beta\omega}{f(\epsilon)}+\ln\left[1-e^{-\beta\left(\frac{\omega}{f(\epsilon)}-\mu\right)}\right]+\ln\left[1-e^{-\beta(\frac{\omega}{f(\epsilon)}+\mu)}\right]\right\}
\end{eqnarray}

Let us highlight here that the above expression for $\ln Z$ must be worked under the consideration of the convergence criteria from the exponentials, i.e., lim$_{\beta\rightarrow \infty} e^{-\beta\left(\frac{\omega}{f(\epsilon)} \pm\mu\right)}<\infty$, from which follows that
\begin{equation}\label{convergence}
    \left|f(\epsilon)\mu(\beta)\right|\leq m.
\end{equation}
This is a modification of the usual convergence condition ($|\mu|\leq m$) very well known in the literature, which was first stated in the work of Haber and Weldon \cite{Haber:1981fg}. As $\beta$ increases, the log contributions in equation $(\ref{generatingfunctional})$ decreases so that for some critical temperature $\beta_c$ defined by $\left|f(\epsilon)\mu(\beta_c)\right|= m$, only the first two terms in the equation~($\ref{generatingfunctional}$) become significant. These two remaining contributions correspond to the zero mode and to the vacuum zero-point respectively.

However, under this same condition ($\left|f(\epsilon)\mu_c\right|= m$) the zero mode term in equation~(\ref{generatingfunctional}) seems to vanish because $f^2(\epsilon)\mu^2-m^2$ goes to zero. In a such scheme there is no charge conservation. Hence, in order to preserve the charge conservation, the U(1) symmetry forces the parameter $\zeta$ to be non zero, so that $(f^2\mu^2-m^2)\zeta^2$ is finite. Therefore, the zero mode becomes macroscopically occupied in the condensed phase.

In addition, the vacuum zero-point contribution to $\ln Z$ is divergent and independent of the occupation number. However, it is not important to the formation of the Bose-Einstein condensate because it can be drop out by some renormalization procedure \cite{kapusta}.

The usual relation
\begin{equation}
    \frac{PV}{T}=\ln Z,
\end{equation}
gives us the equation of state for the system. Hence the pressure is given by
\begin{eqnarray}\label{pressurecomplete}
    \nonumber P&=&(f^2\mu^2-m^2)\zeta^2-\frac{1}{\beta}\int\frac{d^3p}{(2\pi)^3}\left\{\left(\frac{\beta}{f}\right)\omega+\ln\left[1-e^{-\beta\left(\frac{\omega}{f}-\mu\right)}\right]+\ln\left[1-e^{-\beta(\frac{\omega}{f}+\mu)}\right]\right\}
\end{eqnarray}
The internal energy can be written as
\begin{eqnarray}\label{energy}
    \nonumber E&=&-\frac{\partial}{\partial \beta}\ln Z\\
   &=&-V(f^2\mu^2-m^2)\zeta^2-V\int\frac{d^3p}{(2\pi)^3}\left\{-\frac{\omega}{f}+\frac{-\mu-\frac{\omega}{f}}{e^{\beta\left(\mu+\frac{\omega}{f}\right)}-1}-\frac{\mu-\frac{\omega}{f}}{e^{\beta\left(\frac{\omega}{f}-\mu\right)}-1}\right\}.
\end{eqnarray}
The specific heat at constant volume is expressed by
\begin{eqnarray}\label{heat}
    \nonumber C_v&=&\frac{\partial E}{\partial T}\\
    \nonumber&=&-V\beta^2\int\frac{d^3p}{(2\pi)^3}\left\{\frac{\left(\mu-\frac{\omega}{f}\right)\left(\frac{\omega}{f}-\mu\right)e^{-\beta\left(\frac{\omega}{f}-\mu\right)}}{1-e^{-\beta\left(\frac{\omega}{f}-\mu\right)}}+\frac{\left(\mu-\frac{\omega}{f}\right)\left(\frac{\omega}{f}-\mu\right)e^{-2\beta\left(\frac{\omega}{f}-\mu\right)}}{\left[1-e^{-\beta\left(\frac{\omega}{f}-\mu\right)}\right]^2}\right.\\
    &&+\left.\frac{\left(-\mu-\frac{\omega}{f}\right)\left(\frac{\omega}{f}+\mu\right)e^{-\beta\left(\frac{\omega}{f}+\mu\right)}}{1-e^{-\beta\left(\frac{\omega}{f}+\mu\right)}}+\frac{\left(-\mu-\frac{\omega}{f}\right)\left(\frac{\omega}{f}+\mu\right)e^{-2\beta\left(\frac{\omega}{f}+\mu\right)}}{\left[1-e^{-\beta\left(\frac{\omega}{f}+\mu\right)}\right]^2}\right\}
\end{eqnarray}
The charge density is written as
\begin{eqnarray}\label{charge}
    \nonumber\rho&=&\frac{1}{\beta V}\frac{\partial \ln Z}{\partial \mu}\\
     &=& 2f^2\mu\zeta^2+\int\frac{d^3p}{(2\pi)^3}\left[\frac{e^{-\beta  \left(\frac{\omega}{f} -\mu \right)}}{1-e^{-\beta  \left(\frac{\omega}{f} -\mu \right)}}-\frac{e^{-\beta  \left(\mu +\frac{\omega}{f} \right)}}{1-e^{-\beta  \left(\mu +\frac{\omega}{f} \right)}}\right].
\end{eqnarray}

At this point it is important to discuss the role played by the rainbow functions in the critical temperature that sets the BEC. Hence, in what follows we will consider the three mostly adopted cases for the rainbow functions.

\subsection{case I}

Let us consider the case when
\begin{eqnarray}
    f(\epsilon=E/E_P)=1,\,\,\,g(\epsilon)=\sqrt{1-\xi (E/E_P)^s},
\end{eqnarray}
where $E_P$ is the Planck energy, that works as a natural cutoff of the system. Thus, in order to solve the integral in Eq. (\ref{charge}), we will use spherical coordinates and change its integration measure for perform it with respect to $E=\omega$, so that $d^3p=4\pi p^2dp=4\pi p^2(E)(dp/dE) dE$. Following, we expand the integrand around $\xi=0$ up to first order (making $s=1$) and then around $m/T=0$ up to same order. After these steps, we operate the integration from $E=0$ to $E=\infty$ and expand the result for high temperatures. The obtained leading terms are
\begin{equation}\label{rhocaseI}
    \rho\approx\frac{1}{3} m T_c^2 \left(1+\xi\frac{36   \zeta (3)}{\pi ^2 E_P}T_c\right).
\end{equation}
The rainbow-dependent term in the above equation is dimensionless, since the critical temperature is divided by the Planck energy and the $\xi$ parameter is also dimensionless. To obtain this result, we have taken into account that the chemical potential must be equal to the upper limit of the convergence criteria, i.e., $\mu=\pm m$. Moreover, in this case we can see that $\xi=0$ recovers the usual BEC critical temperature\cite{Haber:1981fg}.

A phenomenological analysis can reveal how much of the rainbow gravity can be present and even influence current high-energy physics experiments. In fact, we can infer a upper bound in the factor $\xi T_c/E_P$ given in Eq. (\ref{rhocaseI}) by considering experiments of collisions involving pions, for which the corresponding BEC was reached at a range of critical temperatures of the order of MeV \cite{Begun}. Thus, based on the uncertainty in the current measurements of the charged pion mass given in \cite{PartData}, $|\Delta m_{\pi}/m_{\pi}|\approx 10^{-6}$, we obtain
\begin{equation}\label{bound}
    \xi\frac{T_c}{E_P} \lesssim 10^{-7}.
\end{equation}

\subsection{case II}

Considering the case when
\begin{eqnarray}
    f(\epsilon=E/E_P)=g(\epsilon=E/E_P)=\frac{1}{1-\xi(E/E_P)}.
\end{eqnarray}
For this case, the same procedure employed in the previous case yields the following correction to the critical temperature that sets the BEC,
\begin{eqnarray}\label{rhocaseII}
    \rho\approx\frac{1}{3} m T_c^2\left(1+\xi\frac{54\zeta(3)}{\pi^2 E_p}T_c\right).
\end{eqnarray}

The upper bound in the factor $\xi T_c/E_P$ for the present case keeps unchanged since the slight modification of the equation (\ref{rhocaseII}) in comparison with (\ref{rhocaseI}) does not affect the magnitude order of (\ref{bound}).

\subsection{case III}

Finally let us consider the case when
\begin{eqnarray}
    f(\epsilon=E/E_P)=\frac{e^{\xi(E/E_P)}-1}{\xi(E/E_P)},\,\,\, g(\epsilon=E/E_P)=1
\end{eqnarray}

Similarly, for the BEC to occur we must have the chemical potential equals to the upper limit of the convergence criteria, i.e., $\mu=\pm \frac{m}{f(\epsilon)}$. Setting such condition into  (\ref{charge}) we obtain the corrected critical temperature for the present case as
\begin{eqnarray}\label{rhocaseIII}
    \rho\approx\frac{1}{3} m T_c^2\left(1+\xi\frac{63\zeta(3)}{\pi^2 E_p}T_c\right).
\end{eqnarray}
Also there is no significant modification in the bound presented in (\ref{bound}).

\begin{figure}[h!]
    \centering
    \includegraphics{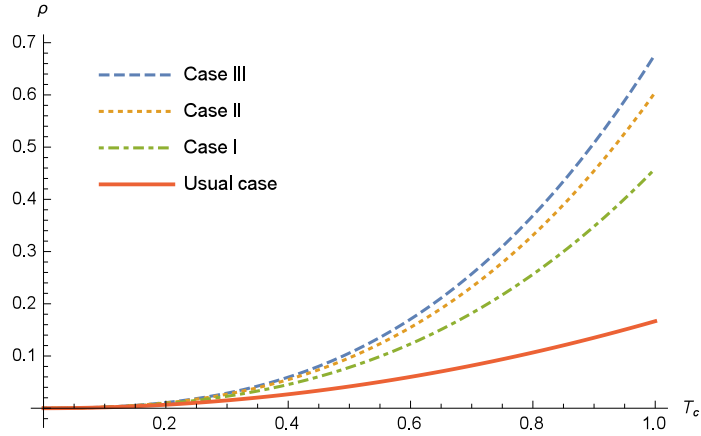}
    \caption{Charge density $\rho$ as a function of $T_c$ for the three cases and the usual relativistic BEC. Here we have considered Planck units in which $E_P=1$ and set $\xi=0.4$ and $m=0.5$}
    \label{fig1}
\end{figure}

All three cases present a very similar behaviour regarding the connection between the charge density and the critical temperature. As we can see clearly from figure (\ref{fig1}), for a given temperature, any of the analyzed cases ($\xi\neq 0$) requires a greater charge density than the usual case ($\xi=0$) in order to establish the relativistic BEC, with the first case requiring the largest density and the third one the lowest. Such a feature can be confirmed by the following discussion.

The fraction of the particles which are not in the 4-D relativistic BEC at an arbitrary temperature, $T<T_c$, is given by
\begin{equation}
    F=\frac{\Delta \overline{N}}{\overline{N}}=1-\frac{T^2}{T_c^2}\left[1-k(T_c-T)\right],
\end{equation}
in first order of $\xi$, with $k\geq0$ corresponding to the factors which multiply the critical temperature within the parenthesis given in Eqs. (\ref{rhocaseI}) (\ref{rhocaseII}), and (\ref{rhocaseIII}). We can notice that such a fraction is greater than the one registered in the usual case ({\it i.e.}, without the influence of the rainbow gravity, $k=0$).

\begin{figure}[h]
\begin{center}
\includegraphics[width=0.5\textwidth]{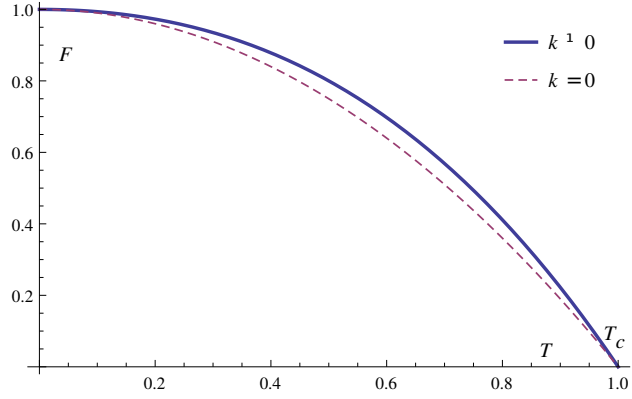}
\caption{Fraction of the particles which are not in the 4-D relativistic BEC, as a function of the temperature, $T\leq T_c$. The parameters settings are $k=0.4$ and $T_c=1$, in Planck units.}\label{FigFrac}
\end{center}
\end{figure}

Fig \ref{FigFrac} reveals how hard it is for the particles to enter the condensed state in the rainbow gravity scenario compared to the usual case. In other words, fewer particles are in that state for a given temperature. Our interpretation is that an upper limit for the energy (the Planck scale) imposes an extra difficulty for one raising the particle energy and thus sharing the condensed state. Notice that when $T=0$ ($T=T_c$), all particles are out of (in) such a state.

\section{Final Remarks}

In this work, we studied the relativistic Bose-Einstein condensate in the context of rainbow gravity. By computing the generating functional we extracted the thermodynamics parameters, such as pressure, energy, specific heat at constant volume and charge density. From the charge density we computed the corrected critical temperature $T_c$ that sets the Bose-Einstein condensate for the three mostly adopted cases (functions) of the rainbow gravity. For each rainbow function we obtained a cubic correction on the critical temperature. We could see in figure (\ref{fig1}) that all three cases present a very similar behaviour regarding the connection between the charge density and the critical temperature. It became clear that for a given temperature, any of the rainbow cases ($\xi\neq 0$) requires a greater charge density in order to establish the relativistic BEC with respect to the case without rainbow's gravity.

Obtaining the corrected critical temperature allowed us to compute an upper bound for a combination of quantities involved in the model from the experiments in high-energy collisions with pions. The upper bound obtained is the same for the three analyzed rainbow functions. Thus, a slight modification in the critical temperature for these considered situations does not affect the order of magnitude of the parameter $\xi$. This result points to the need for even more elaborate experiments and measurements at high energies to obtain tighter bounds on the theory parameters since Planck’s scale is well above the energy scale of the actual experiments.

Finally, we have analyzed the fraction of the particles left out of the relativistic BEC at a temperature lower than the critical one and verified that such a fraction is higher than the one observed in the usual case ({\it i.e.}, without rainbow gravity), which is corroborated by Fig.2. That means that the barrier imposed by the $E_P$ cutoff turns harder these particles to entry the condensate state.

A future perspective for this work consists of analyzing self-interacting scalar fields as well as relativistic BEC in 2+1 dimensions in the rainbow gravity context.

\section*{Acknowledgements}

CRM would like to thank Conselho Nacional de Desenvolvimento Cient\'{i}fico e Tecnol\'{o}gico (CNPq) and Funda\c{c}\~{a}o Cearense de Apoio ao Desenvolvimento Cient\'{\i}fico e Tecnol\'{o}gico (FUNCAP), under grant PRONEM PNE-0112-00085.01.00/16, for the partial financial support.




\begin{thebibliography}{99}

\bibitem{Amelino1} G. Amelino-Camelia, Relativity in space-times with short distance structure governed by an observer independent (Planckian) length scale, Int. J. Mod. Phys. D11 (2002) 35–60,
[gr-qc/0012051].

\bibitem{Amelino2} G. Amelino-Camelia, Doubly Special Relativity, Nature {\bf418}, 34 (2002).

\bibitem{Magueijo} J. Magueijo and L. Smolin, Lorentz invariance with an invariant energy scale, Phys. Rev. Lett. 88 (2002) 190403, [hep-th/0112090].

\bibitem{Galan} Galan and G. A. Mena Marugan, Quantum time uncertainty in a gravity’s rainbow formalism, Phys. Rev. D70 (2004) 124003, [gr-qc/0411089].

\bibitem{Samuel} V. A. Kostelecky and S. Samuel, Spontaneous breaking of Lorentz symmetry in string theory, Phys. Rev.
{\bf D39}, 683 (1989).

\bibitem{Pullin} R. Gambini and J. Pullin, Nonstandard optics from
quantum spacetime, Phys. Rev. {\bf D59}, 124021 (1999).

\bibitem{Carroll} S. M. Carroll, J. A. Harvey, V. A. Kostelecky, C. D. Lane, and T. Okamoto, Noncommutative Field Theory and
Lorentz Violation, Phys.Rev.Lett. {\bf87}, 141601 (2001).

\bibitem{Channuie} A. Chatrabhuti, V. Yingcharoenrat, and P. Channuie, Starobinsky Model in Rainbow Gravity, Phys. Rev. {\bf D93}, 043515 (2016).

\bibitem{Camelia2} G. Amelino-Camelia, J. R. Ellis, N. E. Mavromatos, D. V. Nanopoulos, and S. Sarkar,
Tests of quantum gravity from observations of gamma-ray bursts, Nature 393 (1998) 763–765, [astro-ph/9712103].

\bibitem{Leiva} C. Leiva, J. Saavedra, and J. Villanueva, The Geodesic Structure of the Schwarzschild Black Holes in Gravity’s Rainbow, Mod. Phys. Lett. A24 (2009) 1443–1451, [arXiv:0808.2601].

\bibitem{Li} H. Li, Y. Ling, and X. Han, Modified (A)dS Schwarzschild black holes in Rainbow spacetime, Class. Quant. Grav. 26 (2009) 065004, [arXiv:0809.4819].

\bibitem{Ali} A. F. Ali, Black hole remnant from gravity's rainbow, Phys. Rev. D89 (2014), no. 10 104040,
[arXiv:1402.5320].

\bibitem{Valdir1} V. B. Bezerra, H. R. Christiansen, M. S. Cunha, and C. R. Muniz, Exact solutions and
phenomenological constraints from massive scalars in a gravity's rainbow spacetime, Phys. Rev. D96, 024018, (2017), [arXiv:1704.0121].

\bibitem{Feng} Zhong-Wen Feng and Shu-Zheng Yang, Thermodynamic phase transition of a  hole in rainbow gravity, Phys. Lett. {\bf B772}, 737 (2017).

\bibitem{Li1} P. Li, M. He, J-C. Ding, X-R. Hu, and J-B. Deng, Thermodynamics of charged AdS black holes in rainbow gravity, Adv. High E. Phys. 1043639 (2018).

\bibitem{Ronco} I. P. Lobo and M. Ronco,
Rainbow-Like Black-Hole Metric from Loop Quantum Gravity, Universe  4(12), 139 (2018).

\bibitem{Ednaldo} E. L. B. Junior, M. E. Rodrigues, and M. V. de S. Silva, Regular Black Holes in Rainbow Gravity, Nuc. Phys. {\bf B961}, 115244 (2020).

\bibitem{Dehghani} M.Dehghani, AdS4 black holes with nonlinear source in rainbow gravity, Phys. Lett. {\bf B801}, 135191 (2020).

\bibitem{Valdir3} P. H. Morais, G. V. Silva, J. P. Morais Gra\c{c}a, and V. B. Bezerra, Thermodynamics and remnants of Kiselev black holes in rainbow gravity, Gen. Rel. Grav. {\bf54}, 16 (2022).

\bibitem{Hamil} B. Hamil and B. C. L\"u{}tf\"{u}o\v{g}lu, Effect of Snyder–de Sitter Model on the black hole thermodynamics in the context of rainbow gravity, Int. J. Geom. Meth. Mod. Phys. {\bf19}, 2250047 (2022).

\bibitem{Valdir1.5} V. B. Bezerra, I. P. Lobo, H. F. Mota, and C. R. Muniz, Landau Levels in the Presence of a Cosmic String in Rainbow Gravity, Ann. Phys. {\bf401}, 162 (2019).

\bibitem{Bakke} K. Bakke and H. Mota, Aharonov–Bohm effect for bound states in the cosmic string spacetime in the context of rainbow gravity, Gen. Rel. Grav., {\bf52}, 97 (2020).

\bibitem{Santos} L.C.N.Santos, C.E.Mota, C.C.Barros Jr., L.B.Castro, and V.B.Bezerra, Quantum dynamics of scalar particles in the space–time of a cosmic string in the context of gravity’s rainbow, Ann. Phys. {\bf421}, 168276 (2020).

\bibitem{Awad} A. Awad, A. F. Ali, and B. Majumder, Nonsingular Rainbow Universes, JCAP 1310 (2013) 052, [arXiv:1308.4343].

\bibitem{Majumder} B. Majumder, Quantum Rainbow Cosmological Model With Perfect Fluid, Int. J. Mod. Phys. D22 (2013), no. 13 1350079, [arXiv:1307.5273].

\bibitem{Nozari} M. Khodadi, K. Nozari, and H. R. Sepangi, More on the initial singularity problem in gravity's rainbow cosmology, Gen. Rel. Grav. 48 (2016), no. 12 166, [arXiv:1602.0292].

\bibitem{Valdir2} V. B. Bezerra, H. F. Mota, and C. R. Muniz, Casimir effect in the rainbow Einstein's universe, Eurphys. Lett., 120, 10005 (2017).

\bibitem{Hendi} S. H. Hendi, M. Momennia, B. Eslam Panah, and S. Panahiyan, Nonsingular Universe in Massive Gravity’s Rainbow, Phys. Dark Universe {\bf16},  26 (2017) [arXiv:1705.01099].

\bibitem{Marc} M. Montigny, J. Pinfold, S. Zare, and H. Hassanabadi, Klein–Gordon oscillator in a global monopole space–time with rainbow gravity, Eur.Phys.J.Plus{\bf 137}, 1, 54 (2022).

\bibitem{Takeshi} T. Fukuyama and M. Morikawa, Stagflation: Bose-Einstein condensation in the early universe, Phys. Rev. {\bf D80}, 063520 (2009).

\bibitem{Hermano} H. Velten and E. Wamba, Power spectrum for the Bose–Einstein condensate dark matter, Phys.Lett. {\bf B709}, 1 (2012).

\bibitem{Furtado:2020olp} J.~Furtado, A.~C.~A.~Ramos and J.~F.~Assun\c{c}\~ao, Effects of Lorentz violation in the Bose-Einstein condensation, EPL \textbf{132} no.3, 31001 (2020)

\bibitem{Casana:2011bv} R.~Casana and K.~A.~T.~da Silva, Lorentz-violating effects in the Bose\textendash{}Einstein condensation of an ideal bosonic gas, Mod. Phys. Lett. A \textbf{30} no.07, 1550037 (2015)

\bibitem{Tian:2021mka} Z.~Tian and J.~Du, Probing low-energy Lorentz violation from high-energy modified dispersion in dipolar Bose-Einstein condensates, Phys. Rev. D \textbf{103}, no.8, 085014 (2021)

\bibitem{Furtado:2021tzd} J.~Furtado, J.~F.~Assun\c{c}\~ao and A.~C.~A.~Ramos, Bose-Einstein condensation in Ho\v{r}ava-Lifshitz theory, EPL \textbf{134}, no.1, 11003 (2021)

\bibitem{Takeuchi:2015nga} S.~Takeuchi, Bose\textendash{}Einstein condensation in the Rindler space, Phys. Lett. B \textbf{750}, 209-217 (2015)

\bibitem{Magueijo:2002am} J.~Magueijo and L.~Smolin,``Generalized Lorentz invariance with an invariant energy scale,''
Phys. Rev. D \textbf{67} (2003), 044017 [arXiv:gr-qc/0207085 [gr-qc]].

\bibitem{Magueijo:2002xx} J.~Magueijo and L.~Smolin, ``Gravity's rainbow,'' Class. Quant. Grav. \textbf{21} (2004), 1725-1736
[arXiv:gr-qc/0305055 [gr-qc]].

\bibitem{Awad:2013nxa} A.~Awad, A.~F.~Ali and B.~Majumder, ``Nonsingular Rainbow Universes,'' JCAP \textbf{10} (2013), 052 [arXiv:1308.4343 [gr-qc]].

\bibitem{Ali:2014xqa} A.~F.~Ali, ``Black hole remnant from gravity\textquoteright{}s rainbow,'' Phys. Rev. D \textbf{89} (2014) no.10, 104040 [arXiv:1402.5320 [hep-th]].

\bibitem{Santos:2015sva} G.~Santos, G.~Gubitosi and G.~Amelino-Camelia, On the initial singularity problem in rainbow cosmology, JCAP \textbf{08} (2015), 005 [arXiv:1502.02833 [gr-qc]].

\bibitem{Haber:1981fg} H.~E.~Haber and H.~A.~Weldon, Thermodynamics of an Ultrarelativistic Bose Gas, Phys.\ Rev.\ Lett.\  {\bf 46} 1497 (1981).

\bibitem{kapusta} J. I. Kapusta and C. Gale, Finite-Temperature Field Theory: Principles and Applications, Cambridge University Press, 2ed. (2006).

\bibitem{Begun} V. V. Begun and M. I. Gorenstein, Bose-Einstein Condensation in the Relativistic Pion Gas: Thermodynamic Limit and Finite Size Effects, Phys. Rev. {\bf C 77}, 064903 (2008).

\bibitem{PartData} Particle Data Group, P A Zyla {\it et al.}, Review of Particle Physics, Prog. Th. Exp. Phys., 8, 083C01 (2020).










\end{thebibliography}
\end{document}